\documentclass[12pt]{article}
\usepackage{graphicx}
\usepackage{setspace}
\usepackage{amsmath}
\usepackage{amssymb}
\usepackage{fullpage}
\usepackage{xcolor}
\usepackage{verbatim}
\usepackage{wasysym}
\usepackage{listings}
\usepackage{float}
\usepackage[T1]{fontenc}
\usepackage{geometry}
\usepackage{url}
\usepackage{fancyhdr}
\begin{document}
\title{Prediction Model For Wordle Game Results With High Robustness}
\author{Jiaqi Weng, Chunlin Feng}
\date{March, 2023}
\maketitle
\noindent
In this study, we aim to explore various aspects of Wordle by utilizing data analysis and machine learning techniques. Firstly, we investigate the relationship between the date and the number of reported results. After analyzing the data, we found that the first two months of data were influenced by temporary popularity, and therefore, we only modeled the stable part of the data using the ARIMAX model with coefficient values of 9, 0, and 2, and weekdays/weekends as the exogenous variable. We also observed that there is no explicit relationship between word attributes and the percentage of hard mode results based on the combination of submitted results relationship and hard mode percentage relationship. \\
Next, we utilize the Backpropagation Neural Network model to predict the probability distribution of a given word. We initially faced overfitting problems, which were resolved after feature engineering where we extracted letter frequency, number of unique characters, and part of speech etc. Furthermore, we used the K-means cluster algorithm to categorize numerically the difficulty of a given word, and we found that the algorithm was optimized when there were five clusters based on Silhouette Method. We summarized features for all clusters and found several correlations with word attributes. 
Our models indicate that on March 1st, 2023, there will be 12884 submitted results. We also found that the word ``eerie'' has an average of 4.8 tries among all players and is categorized as level 5 of difficulty with our K-means cluster algorithm, which is the hardest level. Additionally, we discovered some interesting features, such as the percentage of loyal Wordle fans and the proportion of them that are willing to take challenges every day. \\
We conducted a series of sensitivity analyses that demonstrated the robustness of our models using various statistical tests, including ADF, ACF, PACF, and cross-validation. In particular, we tested the robustness of our ARIMAX model by providing it with a different dataset due to an unexpected increase in player numbers, and the predicted results still tended to remain around 13,000. \\
Overall, our study offers a novel and effective approach to anticipating the outcome of a Wordle game given a date or a random five-letter word. In the end, we included an letter written to Puzzle Editor of New York Times with our results summarized. 
\newpage

\tableofcontents
\newpage
\section{Introduction}
\subsection{Background}
Wordle is a popular online game where players have to guess a five-letter word chosen by the game. Each time a player submits a word, the game provides feedback on how many letters are correct and in the correct position. If the letter is correct but in the wrong position, it is shown in yellow. If the letter is correct and in the right position, it is shown in green. The goal of the game is to guess the word in as few attempts as possible. \\ \\
The game also has a hard mode, which means once a player has found a correct letter in a word (either yellow or green), those letters must be used in subsequent guesses.  \\ \\
Wordle has become increasingly popular on social media platforms and has gained a significant following. Many players have been trying to develop strategies and mathematical models to anticipate the results of the game and to improve their chances of winning. In this report, we will present a mathematical model for Wordle, which aims to predict the most likely word based on the game's feedback.
\subsection{Problem Statement}
In this contest, we were being asked to develop a model to find out a relationship between the date and the number of results. We also need to predict the distribution of the reported results given a random 5-letter word. Following a comprehensive analysis and research of the problem's background, we have identified that our article should encompass the following aspects:
\begin{itemize}
    \item Develop a model that predicts the number of result submitted on March 3rd, 2003
    \item Develop another model that predicts the distribution of results based on a given word
    \item Develop another model that returns a level of difficulty based on a given word
    \item Demonstrate with evidence that our model offers the most effective approach.
    \item Assess the level of sensitivity of these models
    \item List and describe some other interesting features of this data set.

\end{itemize}

\subsection{Assumptions}
We did have to clean up some data. We searched up the date with `Wordle of the Day' followed by the date and used Google, and the update on the MCM / ICM Website. For the website, we used Rock, Paper, Shotgun which is in our references \cite{WORDLE_ANSWER}.
\begin{table}[H]
    \centering
    \begin{tabular}{|c|c|}
        \hline
         Former Word & Edited Word \\
         \hline\hline
         clen & clean \\
         \hline
         rprobe & probe \\
         \hline
         tash & trash \\
         \hline
         na\"ive & naive\\
         \hline
    \end{tabular}
    \caption{This table shows the former word and the edited word.}
    
    \label{tab:my_label}
\end{table}
\noindent
Note that we also change the \"i to the English `i' in order to make our model. Also, an outlier in the given data with only 4 numerical digits was simply replaced by the average of its left and right values.
\subsection{Definitions and Abbreviations}
\begin{itemize}
    \item ARIMA stands for Autoregressive integrated moving average
    \item ARIMAX stands for Autoregressive Integrated Moving Average with Explanatory Variable
    \item BP stands for Backpropagation
    \item ADF stands for Augmented Dickey-Fuller
    \item A miss indicates not being able to solve a Wordle problem
    \item Consecutive Uncommon Letters refers to letters that may trip people up if those letters are together
    \item ACF stands for autocorrelation function 
    \item PACF stands for partial autocorrelation function
    \item CNN stands for convolutional neural network
    \item BPNN stands for Backpropagation neural network
    \item LSTM stands for Long short-term memory
    \item Dates are put in the form MM/DD or MM/DD/YYYY
\end{itemize}
\section{The Models}
\subsection{Prediction Model For Date vs. Number of Results}

\begin{center}
\includegraphics[scale = 0.4]{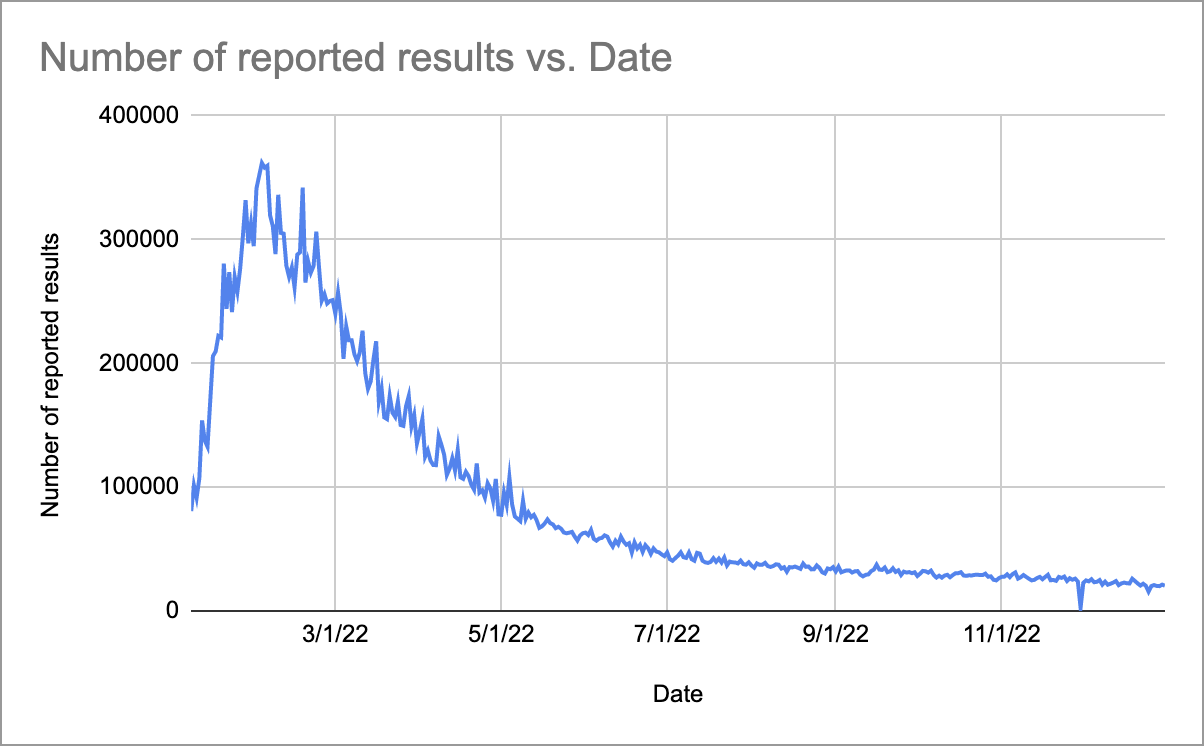}\\
Figure 1
\end{center}
Figure 1 shows the change of number of submitted results from 1/26/2022 to 12/31/2022 \cite{WORDLE_ANSWER}. Before any further actions, we conducted some basic level of data cleaning as explained in Section 2.3. After that, as evidenced by the graphical representation above, the quantity of players demonstrates a gradual rise over time, followed by a decline by February 2$^{\text{nd}}$. We made an assumption that the sudden rise of the results from 1/20 to 2/2 was mostly due to a temporary interest due to publicity. Therefore, we will only consider the decreasing period in our following modelling process. We could easily conduct a linear or quadratic approach for the increasing period, but it does not help at all for us to determine the number of submitted results on 3/1/2023. 
\subsubsection{Model Selection}
The methodology employed in selecting this model was relatively straightforward. Since the given data set contains a total number of 359 days, we divided them into two parts. The first 328 days is the training group, and the last 32 days is the testing group. We have built different models based on the data in the training group, then we tested their accuracy in the testing group. \\
As a model that is needed for time series forecasting, ARIMA seems to be a good choice typically. Although the data trending varies significantly between January to March and March to December. But as Figure 1 demonstrated in the previous section, the number of results seems to become stationary towards the end of the year, which means we could implement ARIMA model only on the second half of the data.  \\
It is also possible to employ an exponential decay model, as the analysis of the data set indicates that the submitted results exhibit a stabilization around 20,000. While this approach is relatively straightforward and adaptable, it is particularly susceptible to the influence of outliers. \\
Therefore, we applied both exponential decay and ARIMA model to approach the problem. As mentioned in section 3.1, we will only be modelling the declining portion of the graph, in other words, our training group will starts on 2/2/2022 and ends on 11/31/2022. Then we will measure the error on the testing group, which is from 12/1/2022 to 12/31/2022. Below is the graph for the exponential decay formula that was generated by MATLAB.
\begin{center}
    \includegraphics[scale = 0.5]{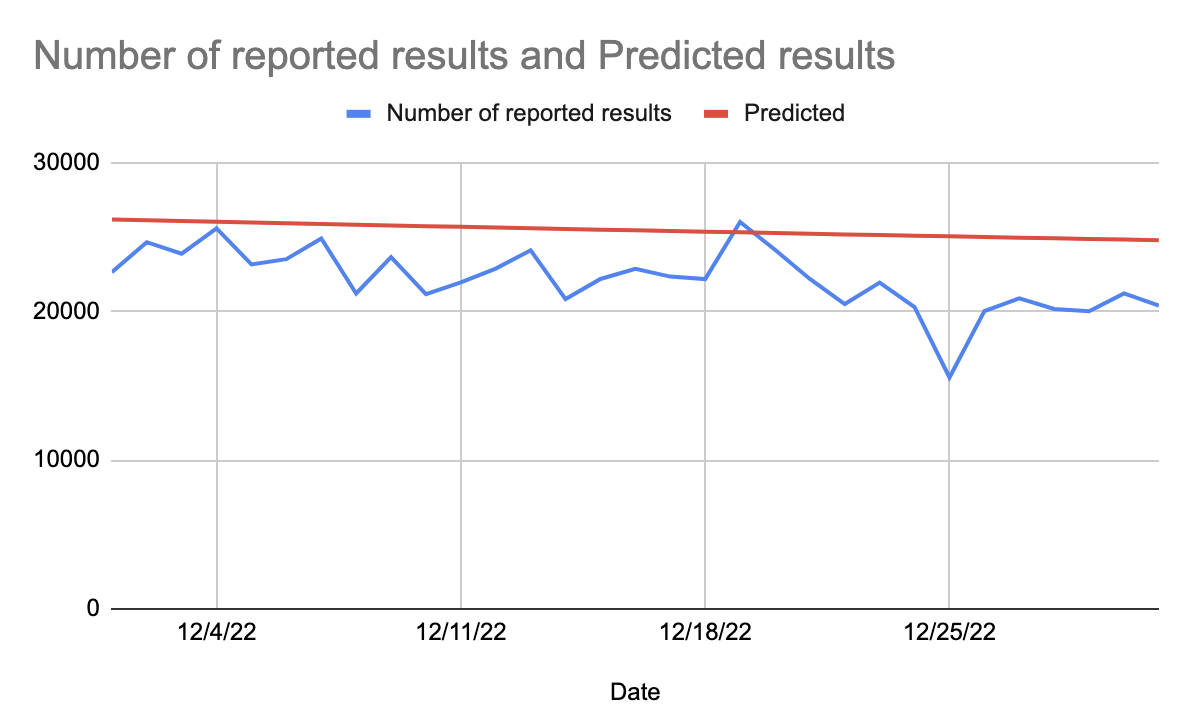} \\
    Figure 2: Curve Fitting of Exponential Decay
\end{center}
As one can see, the exponential decay curve obeys the trend of our testing group. However, it lies above the actual data. On the other hand, we also tested the prediction using standard ARIMA model. By manually adjusting the coefficient, we found that ARIMA(9, 2, 2) fits the testing group well, and the difference graph is shown below:
\begin{center}
    \includegraphics[scale = 0.63]{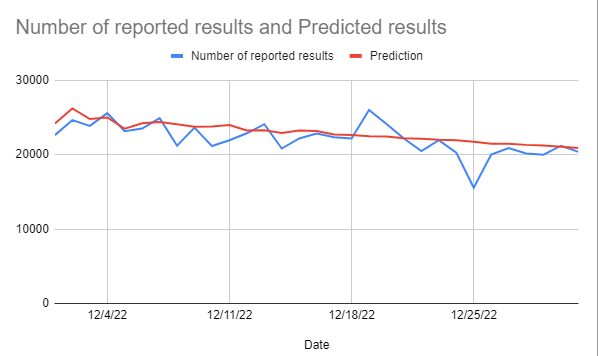}
    \\Figure 3: Curve Fitting of ARIMA(9,2,2)
\end{center}
If the actual number of submitted results on day $k$ is $n_k$ while our predicted number on day $k$ is $n'_k$, then the errors and residuals $w$ can be calculated as $$w = \frac{|n'_k - n_k|}{n_k}$$By calculating the average error of the two models we used, we found that the error for exponential decay is 0.16, while the error for ARIMA model is 0.06. Therefore, we conclude that ARIMA is better than a simple exponential curve fitting at this Wordle game prediction.
\subsubsection{The ARIMA Model}
In time series analysis, ARIMA (Autoregressive Integrated Moving Average) model is one of the most commonly used models for forecasting. ARIMA models are used to model stationary time series data, which have a constant mean and variance over time. The model consists of three components: autoregression (AR), differencing (I), and moving average (MA). Autoregression represents the correlation between the observations and their lagged values, differencing is used to make the time series stationary, and moving average is used to model the noise in the data. The general formula for ARIMA($p$, $d$, $q$) model can be written as follows:

\begin{equation}(1-\sum_{i=1}^p \phi_i L^i)(1-L)^d X_t = (1+\sum_{j=1}^q \theta_j L^j) \epsilon_t \tag{From Reference \cite{PREDICT_STOCK_PRICES_WITH_ARIMA_AND_LSTM}}\end{equation}
where $X_t$ is the time series data, $\phi$ and $\theta$ are the autoregressive and moving average coefficients respectively, $L$ is the lag operator, $d$ is the order of differencing, and $\epsilon_t$ is the white noise error term. The parameters $p$, $d$, and $q$ are usually determined by analyzing the autocorrelation and partial autocorrelation functions of the data. The ARIMA model is a powerful tool for forecasting time series data, and its effectiveness is demonstrated through numerous empirical studies. \\
In the preliminary analysis, the ARIMA (9, 2, 2) model was provisionally selected as a fitting candidate for the curve. Notwithstanding, this choice was merely based on intuitive grounds and thus demands empirical validation. Accordingly, in the subsequent sections, we carried out several diagnostic tests to establish the optimal parameterization of the ARIMA model \cite{ANALYTICS_INDIA_MAG}. \\
First of all, the integrated item, $d$, can be tested by ADF test (as seen in Reference \cite{ANALYTICS_INDIA_MAG}), which tests for the presence of a unit root in a time series. After running through the ADF test in MATLAB, we found that for zero-order difference, the h-value is 1 and the p-value already equals 0.01, which is less than 0.05. This means that our time-series data is already stationary enough without further differentiating, so our d-value should be 0. An underlying premise we adopted for the efficacy of ARIMA(9, 2, 2) during our testing phase was that the time series had already attained a sufficient level of stationarity. Consequently, any additional differentiation is unlikely to significantly impact the predictive outcomes. \\
Next, the autocorrelation function (ACF) and partial autocorrelation function (PACF) serve as fundamental tools for determining the coefficients and orders of the ARIMA model with parameters $q$ and $p$ \cite{ANALYTICS_INDIA_MAG}. Below are the images we got for PACF test which ran through the MATLAB:
\begin{center}
    \includegraphics[scale = 0.5]{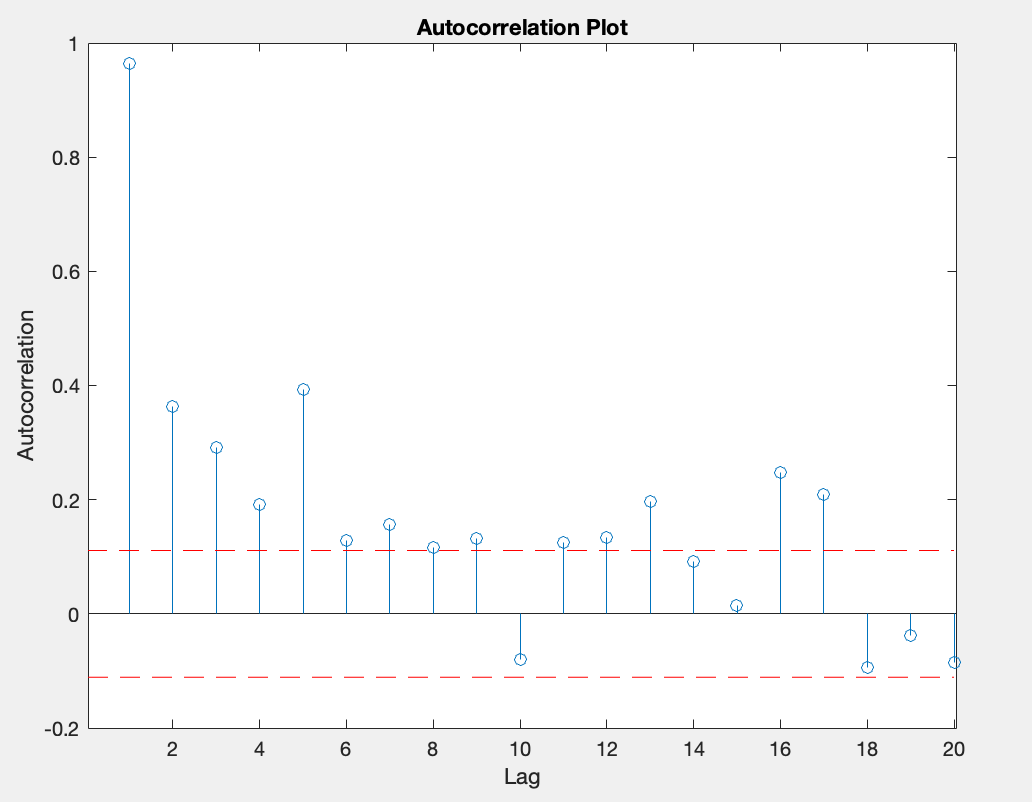}\\
    Figure 4: PACF Test\\
    \includegraphics[scale = 0.5]{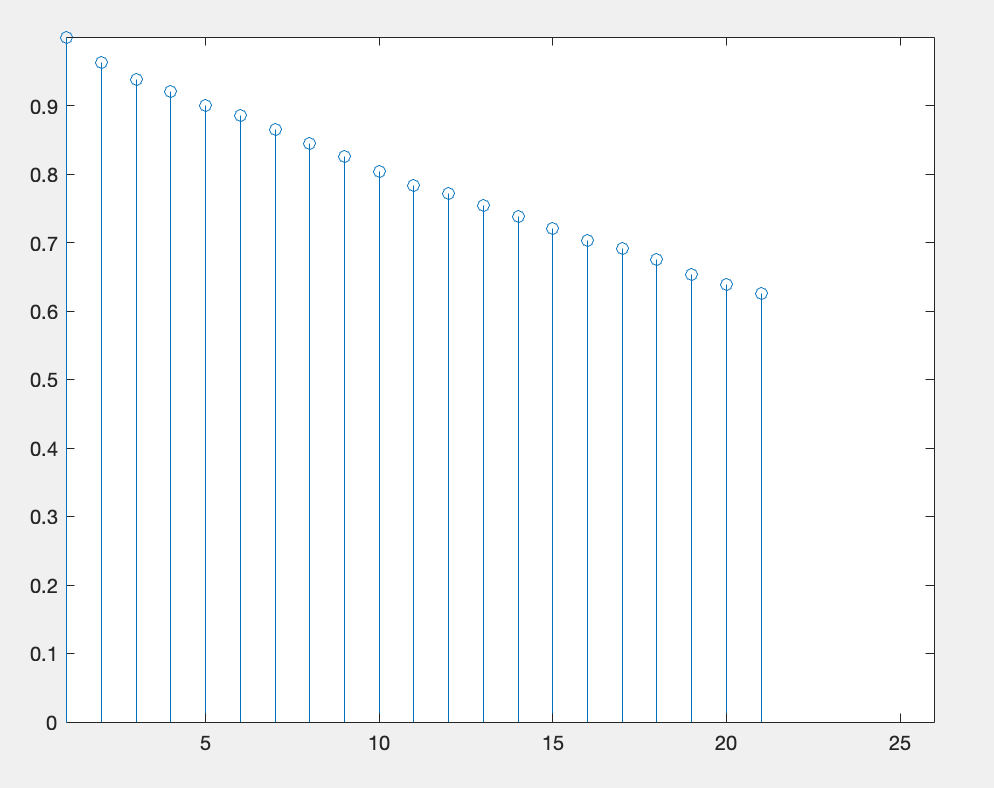}\\
    Figure 5: ACF Test
\end{center}
For the PACF test, the result was fairly elementary. We can see that it is not until the 10$^{\text{th}}$ log when the partial autocorrelation falls into our confidence interval, which was calculated by the formula $CI = \bar{x} \pm z \frac{s}{\sqrt{n}}$. Therefore, our p-value will be 9 since the log counts starts from 0, which surprisingly matches our hypothesis. \\
However, the ACF test, which determines parameter $q$, is way far off. Based on the graph we can see that non of the 20 points lie in the confidence interval. According to [3], the source indicates that our time series is very stationary, so it is possible for us to just choose a small number between below 2. After our manual test on ARIMA(9, 0, 0), ARIMA(9, 0, 1) and ARIMA(9, 0, 2), we found that ARIMA(9, 0, 2) fits the testing group the best. Therefore, our final model will be ARIMA(9, 0, 2), which has the formula $$(1-\sum_{i=1}^9 \phi_i L^i) X_t = (1+\sum_{j=1}^2 \theta_j L^j) \epsilon_t $$
where:
\begin{itemize}
    \item $X_t$ is the value of the time series at time $t$
    \item $L$ is the lag operator and $L^i$ indicates lagging the variable by i times period.
    \item $\phi_1$ to $\phi_9$ are the autoregressive coefficients ($\forall\phi_i$)
    \item $\theta_1$ and $\theta_2$ are the moving average coefficients ($\forall\theta_j$)
    \item $\epsilon_t$ is the error term at time $t$
\end{itemize}
In this model, the time series is differenced zero times ($d = 0$), indicating that it is stationary. The order of the autoregressive component is $p = 9$, meaning that the current value of the time series depends on the previous 9 values. The order of the moving average component is $q = 2$, indicating that the current error term depends on the two previous error terms. \\ \\
To demonstrate the absence of autocorrelation in the residuals of the model and establish its adherence to the assumptions of our ARIMA model, an autocorrelation analysis was conducted and presented in the figure below. The analysis revealed that the residuals of the predicted model result conform to a white noise sequence, thus validating the adequacy of the ARIMA model.
\subsubsection{The Improved ARIMA Model: ARIMAX}
Despite yielding a relatively low error of 0.06 in matching the testing group (12/1 to 12/31), our original ARIMA model remains susceptible to the occurrence of large errors. Additionally, the model did not account for other factors that may impact the equation, such as the attributes of the word on a given day or whether it falls on a weekend or weekday. Therefore, our group conducted another model based on our current ARIMA model, but with other exogenous variables. \\
The first factor we considered was if the given date lies between weekdays or weekends. We use Google Sheets to run though the whole data set and then separate and categorize regarding with weekdays and ends. Below is the data we summarized:
\begin{center}
    \includegraphics[scale = 0.5]{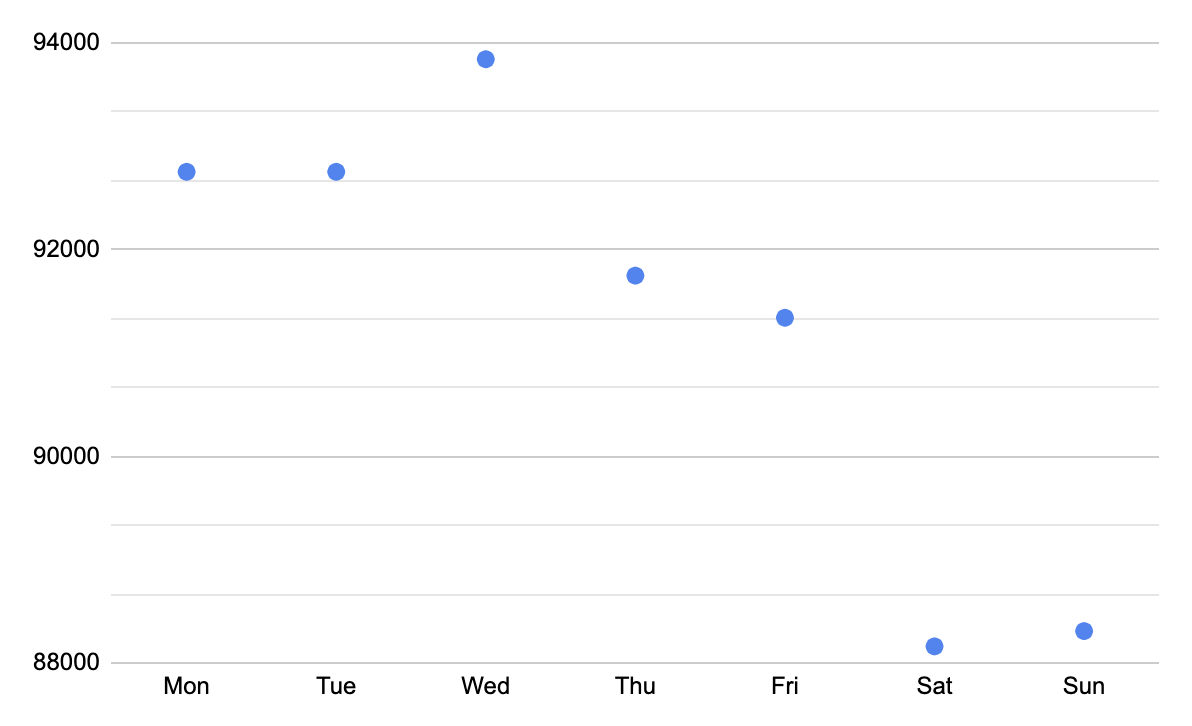}\\Figure 6: Average Number of Results Catagorized by Weekdays
\end{center}
From the observation, we found that it does have a certain level of negative correlation regarding with either weekends or weekdays. Therefore, we summarized this relationship with the formula $$W = \frac{\overline{X_{weekdays}} - \overline{X_{weekends}}}{\overline{X_{weekdays}}}$$where $\overline{X}$ is the mean value of either weekdays or weekends. The relationship we calculated equals to $0.04599$, meaning that in general, the reported results on weekends is $4.59 \%$ smaller than the ones on weekdays. \\
Therefore, our exogenous variable, $\beta W_t$, will be an indicator representing either the weekdays/weekends indicator. It is defined as follows : 

$$W_t = \begin{cases} 1 & \text{(if day $t$ is a weekday)} \\  1 - 0.0459 & \text{(if day $t$ is a weekend)} \end{cases}$$
As $W_t$ behaves more akin to a modifier that influences the outcome of the aggregate $X_t$ rather than a distinct variable, we multiplied it into the left-hand side. By substituting this definition into the initial ARIMA equation, we formulated our first ARIMAX model, presented below:
$$(1-\sum_{i=1}^9 \phi_i L^i)W_t X_t = (1+\sum_{j=1}^2 \theta_j L^j) \epsilon_t,  W_t = \begin{cases} 1, & \text{(if day $t$ is a weekday)} \\ 1 - 0.0459 & \text{(if day $t$ is a weekend)} \end{cases}$$
Therefore, this will be our new ARIMAX formula. After that, we conducted another test using our new model ARIMAX(9, 0, 2) into our training group, and then test its accuracy in our testing group. As a result, the average error rate dropped down by 0.002.
\begin{center}
    \includegraphics[scale = 0.45]{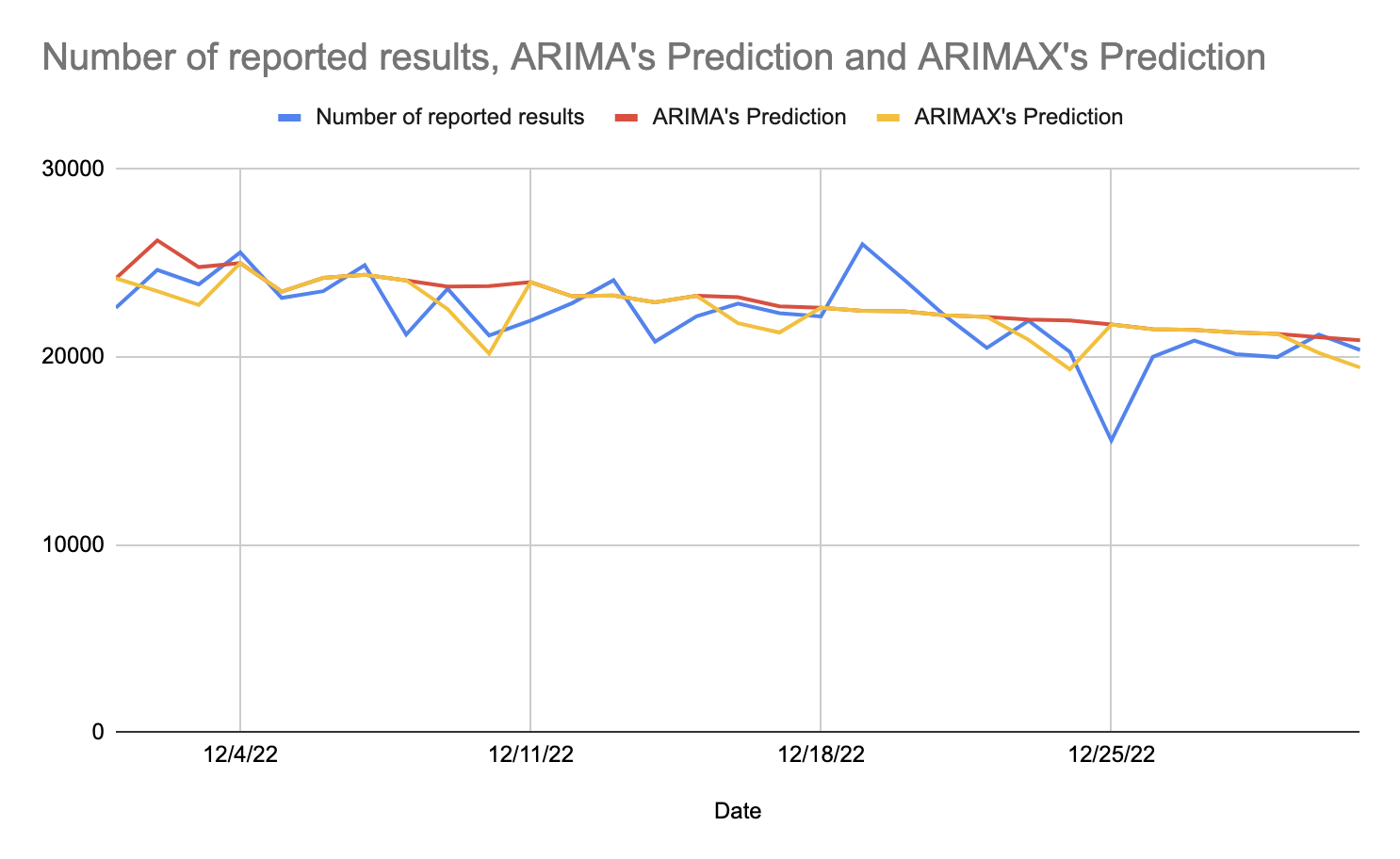}\\Figure 7: Comparison Between ARIMA and ARIMAX
\end{center}
\subsection{Predicting Percentage of Hard Mode Percentage For Next Words}
Before any further movement is done, we plotted the diagram regarding with the date and the percentage of hard mode, as shown below:
\begin{center}
    \includegraphics[scale = 0.35]{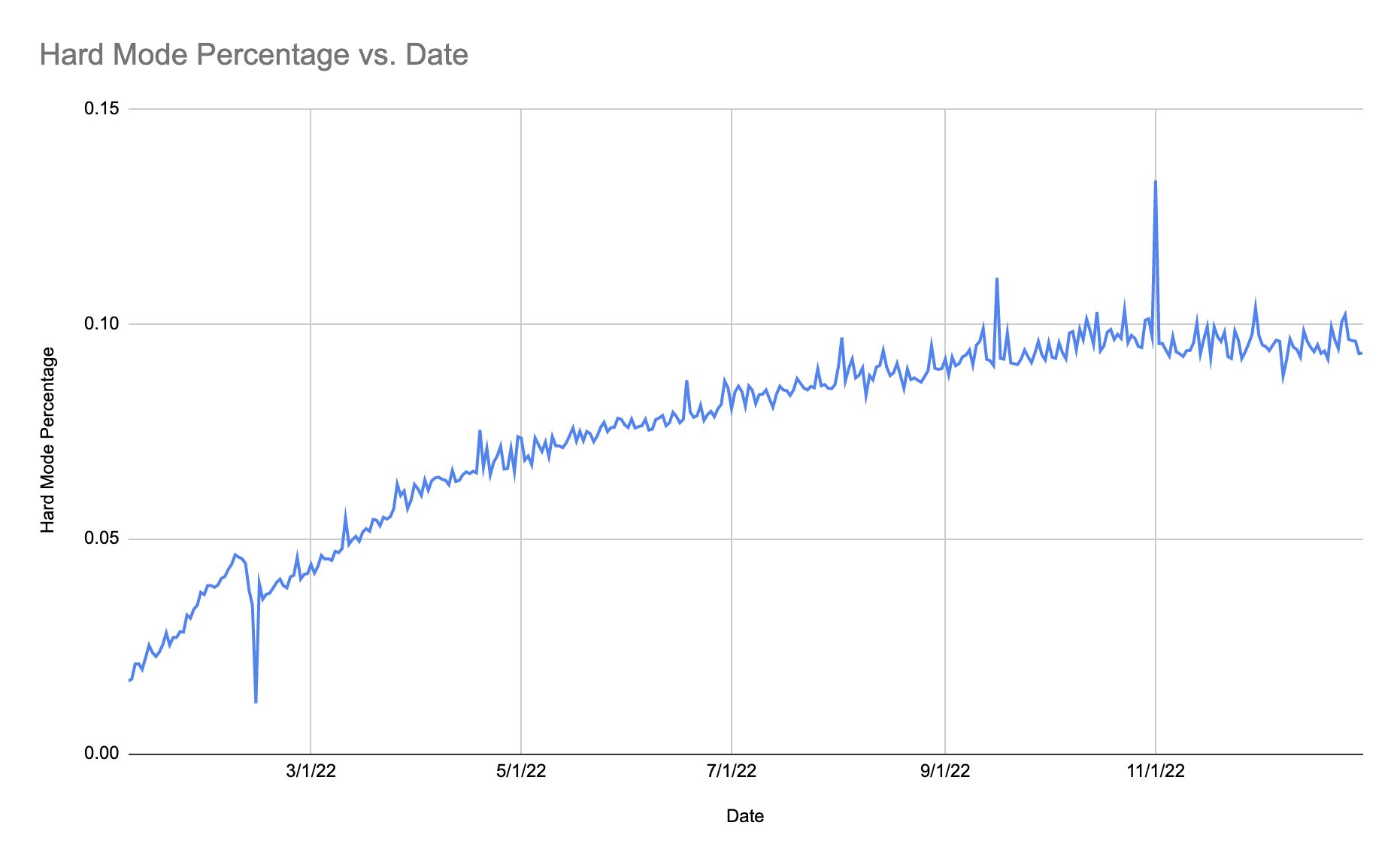}\\Figure 8: Hard Mode Percentage vs. Date
\end{center}
Overall, the trends for percentage of hard mode shows an upward tendency. After 9/1 of 2022, this percentage tends to stay in $8 \%$ with plus or minus $1 \%$ despite some outliers. Based of the graph from our first glance, it can be inferred that there is no explicit relationship between the attributes of a word and the hard mode percentage. This study aims to investigate the potential correlation between the percentage of difficult mode and the number trend of 2022 in the game of Wordle. Specifically, we propose that the rapid growth in the first two months of 2022 was driven by the promotion by the New York Times, during which new players continuously joined but were unlikely to use hard mode while playing. Following the promotion, we hypothesize that the remaining players are all core players of Wordle, with $8 \% $ of them still willing to try hard mode regardless of word difficulty. We contend that a $1 \%$ fluctuation cannot reflect the influence of word attributes on the hard mode percentage. For example, on November 1st, we identified ``piney'' as a difficult word due to its low frequency of appearance and the presence of the uncommon letter ``y''. However, its hard mode percentage was the highest at $13.33 \%$ among the words played that day. In contrast, some very common words, such as ``drive'', had a hard mode percentage of only $8 \%.$ Therefore, we speculate that in the late stage of the game when the number of players gradually stabilizes, there is no clear relationship between word attributes or difficulty and the hard mode percentage. The subsequent sections will elaborate on our methods for evaluating word difficulty and provide further analysis of the observed patterns.

\subsection{Model for Predicting Percentages For Next Words}
\subsubsection{What We Observed}
As for the model for forecasting the probability distribution of a given word, we initially hypothesized a negative correlation between the level of ``difficulty'' of a word and its corresponding hard mode percentage. To develop a metric for assessing the difficulty of a word, we identified several attributes, such as letter frequency, repetition of letters, and proportion of vowels etc. based on our experience on playing Wordle. First, we realized that double letters (letters that appeared twice), triple letters (letters that appeared three times), consecutive uncommon letters in the third and fourth positions, and uncommon words increased the difficulty for users (caused for there to be more misses percentage wise). The consecutive uncommon letters included `tc' and `dg'. These consecutive letters seemed to make it more difficult for a user to guess the word. We tested on the letter `x' as we believed that that is the most uncommon letter in English, as shown below:
\begin{center}
    \includegraphics[width=0.4\textwidth]{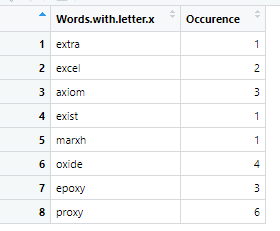}
    \\Table 2: words with the letter `x' (the occurrence refers to the percentage of misses)
\end{center}
We also think words that were more mathematical seemed easier to guess: these words include words such as `axiom', `third', `prime', and `equal'. 
\subsubsection{Back Propagation Neural Network}
Due to the complexity of measuring a word with mathematical formulas, our team chose to use a machine learning model to approach this problem. CNN have been extensively used in image processing tasks; nevertheless, they are restricted in their capability to handle non-image data types. Although LSTM networks have been employed in previous research to predict time-series models, they necessitate a significant amount of data. As a result, in this study, we opted to employ BPNN for our analysis. BPNN is a versatile model capable of handling a wide range of input data types, particularly advantageous for smaller datasets, and proficient in modeling nonlinear relationships between inputs and outputs \cite{youtube}. Our data set consisted of 359 days of Wordle results, making BPNN a suitable choice for our research. \\\\
The first step in our methodology was feature engineering \cite{PATTERN_RECOGNITION_AND_MACHINE_LEARING}. We wish to extract the attributes of each word and measure them numerically so it suits for machine learning. The first approach we adopted was using MATLAB to calculate the frequency of each letter appeared in all 359 given words. This is because as mentioned in the previous section, words with low frequency letters such as  ``x''  shows the increased number of trials to solve. The pie chart is shown below:
\begin{center}
    \includegraphics[scale = 0.5]{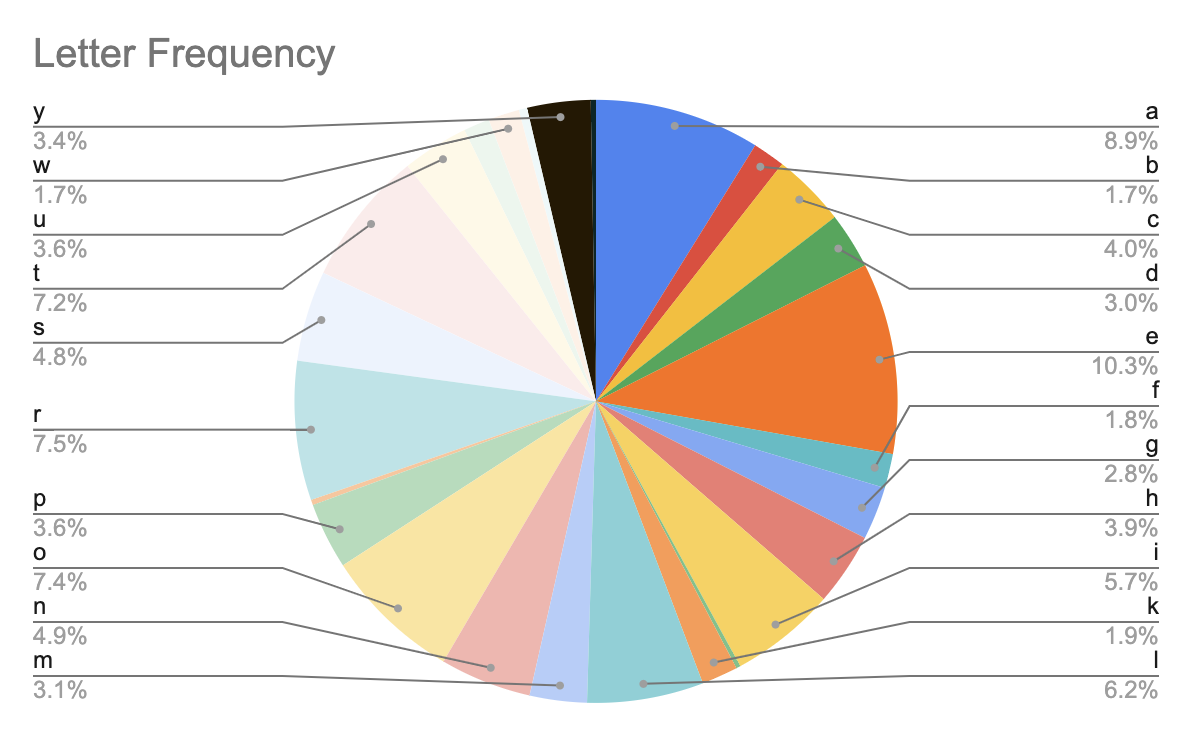}\\Figure 9: Letter Frequency
\end{center}
Furthermore, we agreed that the repetition of letters can also impact the Wordle game's outcome. This is due to the fact that, given a fixed number of unused words, guessing a word with all unique letters is more challenging than guessing a word with duplicated letters. For example, if a player has a letter pool of 7 letters left, and we set aside the validity of the word, the chance of him or her guessing a correct word with 5 distinct letter is: $$\frac{1}{7 \cdot 6 \cdot 5 \cdot 4 \cdot 3} = \frac{1}{2520} $$On the other hand, if the correct word has three repetitive letters, for example, eerie, then the probability of the player getting this word is only $$\frac{1}{7^3 \cdot 6 \cdot 5} = \frac{1}{10920}$$Therefore, we created another list indicating the number of distinct letters. Furthermore, we summarized a table shown in figure 10 below with the first 20 rows:
\begin{center}
    \includegraphics[scale = 0.2]{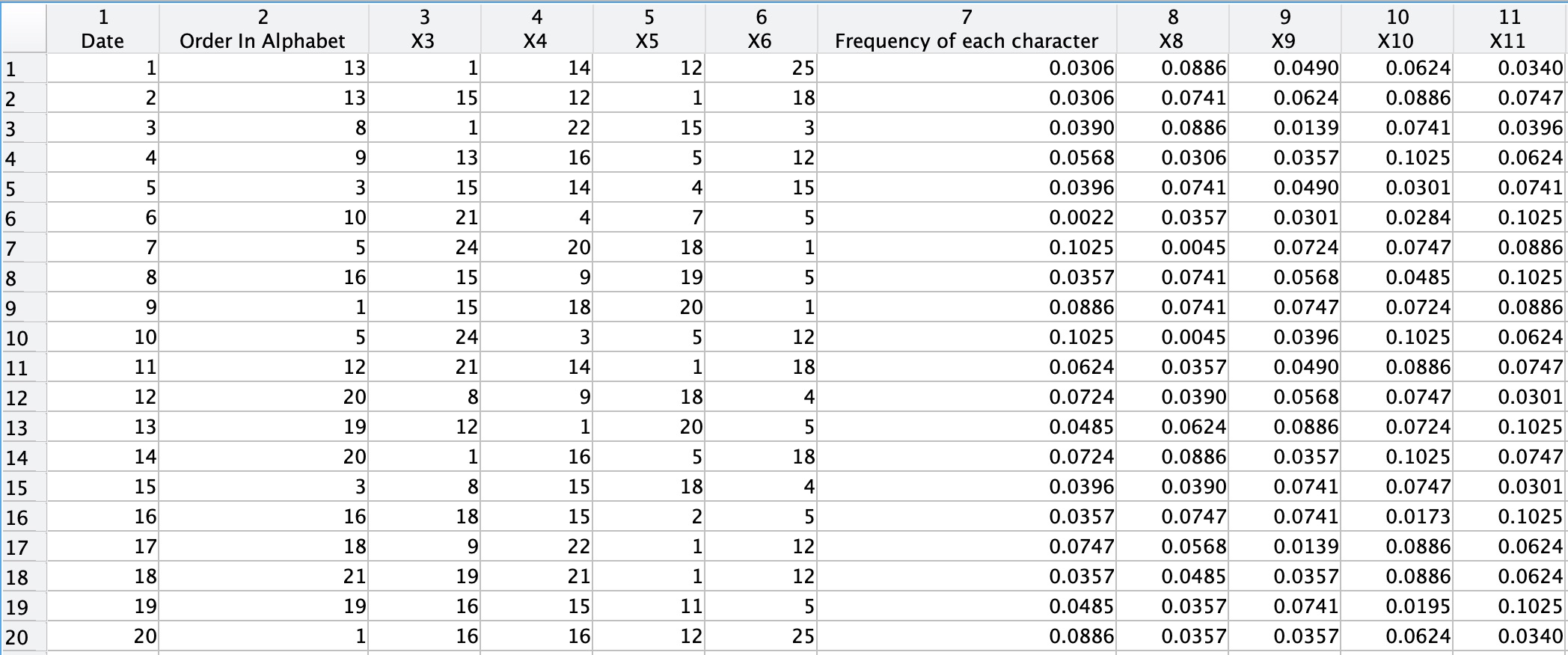}\\Table 3: Input Data
\end{center}
The first column indicates the date of the given word in integer. We assumed the starting date of our model as day 1 (Therefore, the requirement which asks about March 1st 2023 will be integer 419). Then the next 5 columns are the integer representation of the 5 letters in the given word. The next 5 columns are the frequency representation.\\
Using the same methodology in section 2, we also split our dataset into two groups. The first 300 days is the training group, and the last 59 days is the testing group. Similarly, we used the training group to train our model, and test its effeteness on our testing group. \\
The preliminary experiments were unsuccessful due to overfitting, which is a common challenge for nascent machine learning models \cite{PATTERN_RECOGNITION_AND_MACHINE_LEARING}. A comparative analysis of one of the trial outcomes with the original dataset is presented below:
\begin{center}
    \includegraphics[scale = 0.20]{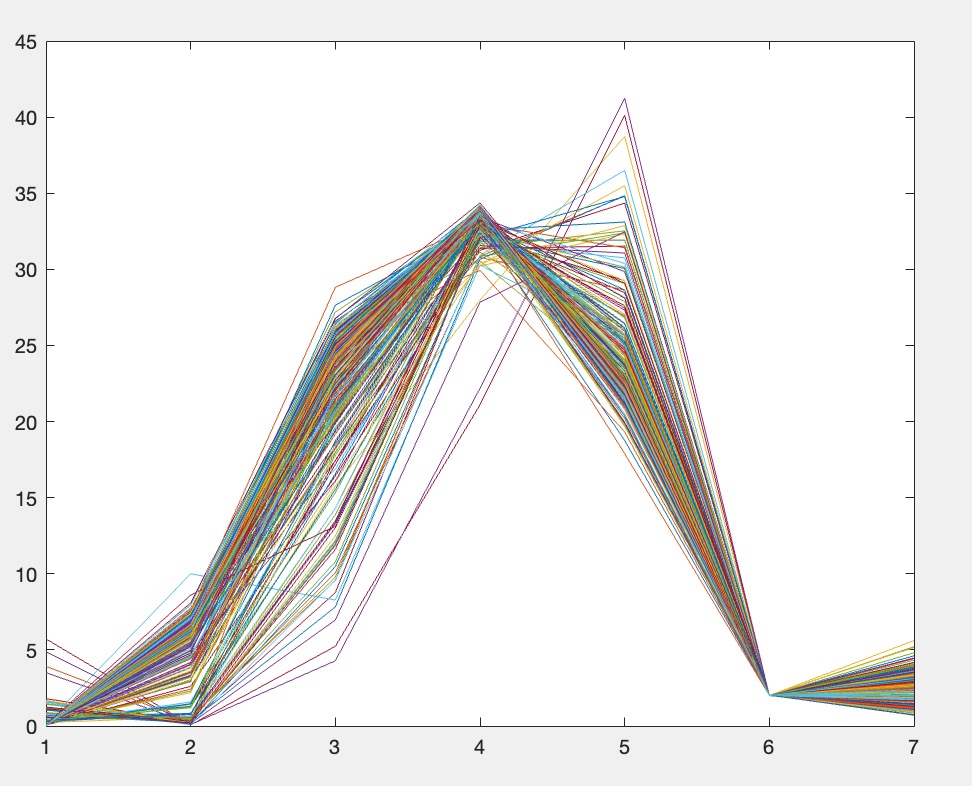}
    \includegraphics[scale = 0.20]{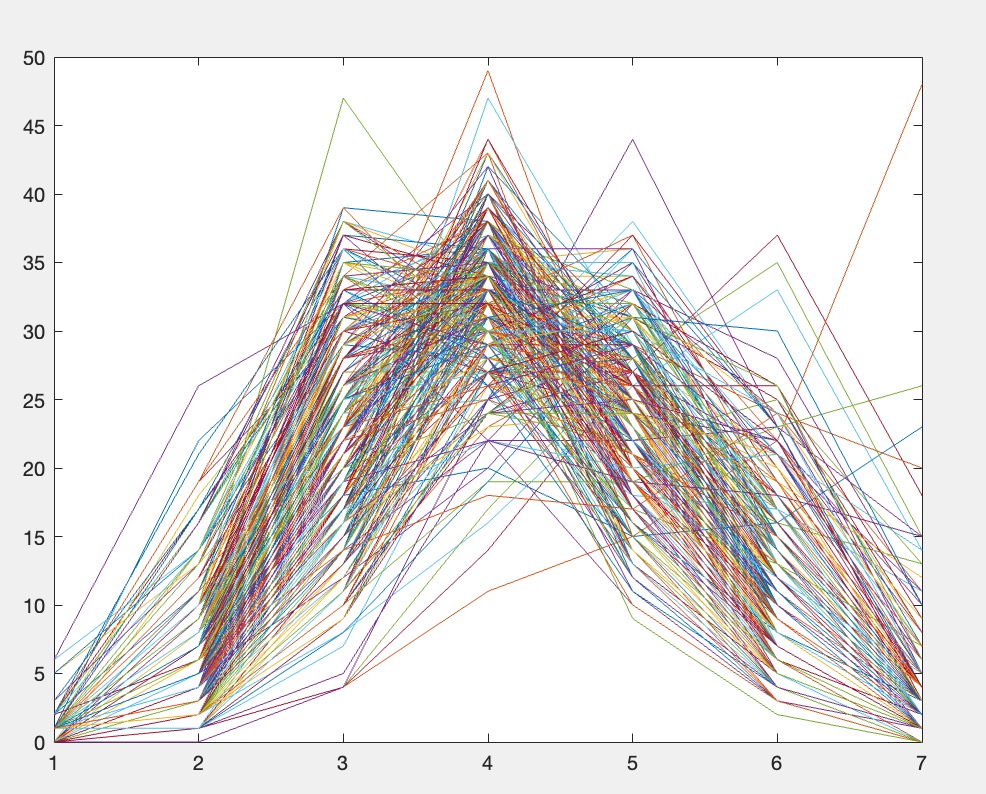}\\Figure 10: Comparison of Predicted and Actual Results
\end{center}
The visualization reveals that our predicted training outcomes exhibit a tendency to converge towards a single point. Additionally, our training process consistently terminated at a significantly lower epoch count of 30-50, rather than the desired 1,000 epochs, which is one of the classical issues of overfitting. The root cause of this phenomenon can be attributed to either the dataset quality or the stochastic nature of the input. Since the dataset is already given and we cannot attain more information, the only thing we could do is specify our feature engineering process. In other words, we need to summarize more features from each given word. \\
We summarized and listed all potential features of a word that could be a factor of the difficulty, or influential to our results, including number of vowels and consonants, number of unique characters, and the part of speech of the word. A table of additional features added is included below.\\
\begin{center}
    \includegraphics[scale = 0.2]{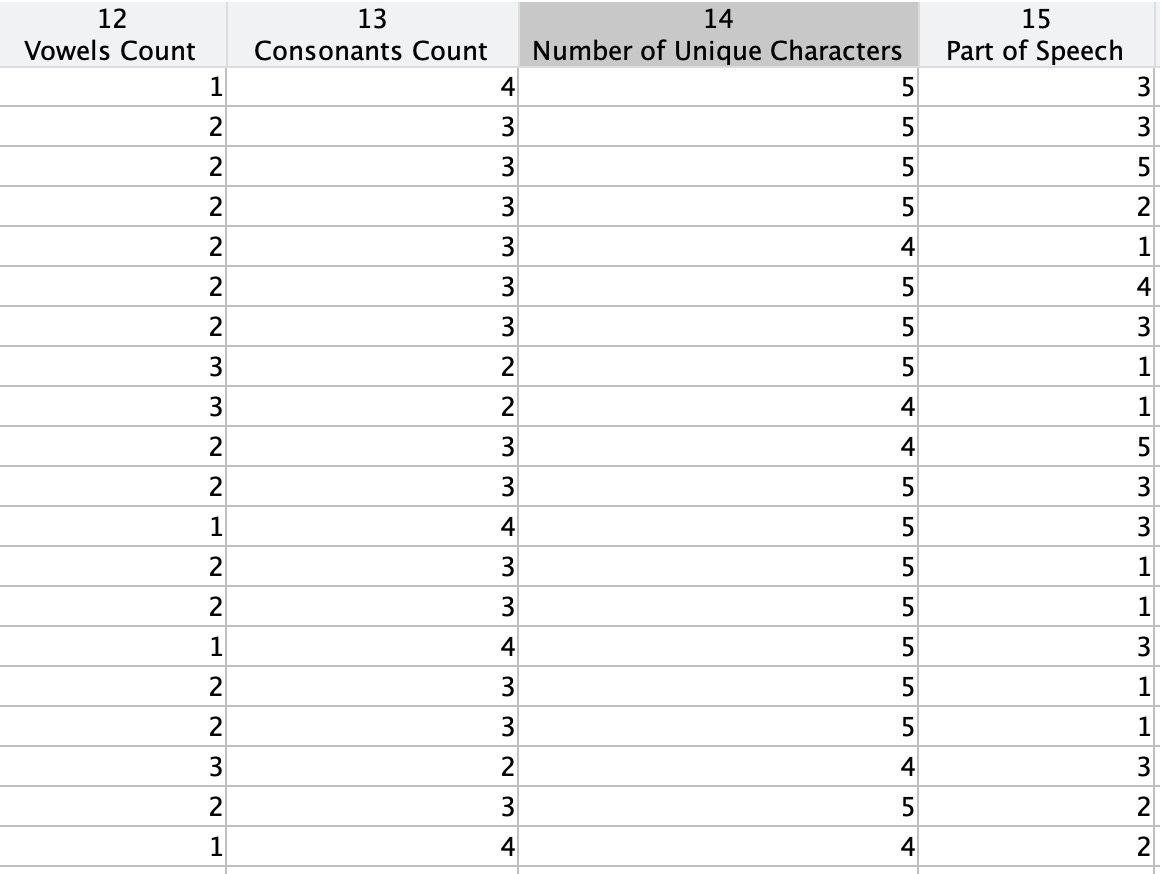}\\Table 4: More Word Features
\end{center}
After that, our new results with regards to the given data is presented below:
\begin{center}
    \includegraphics[scale = 0.20]{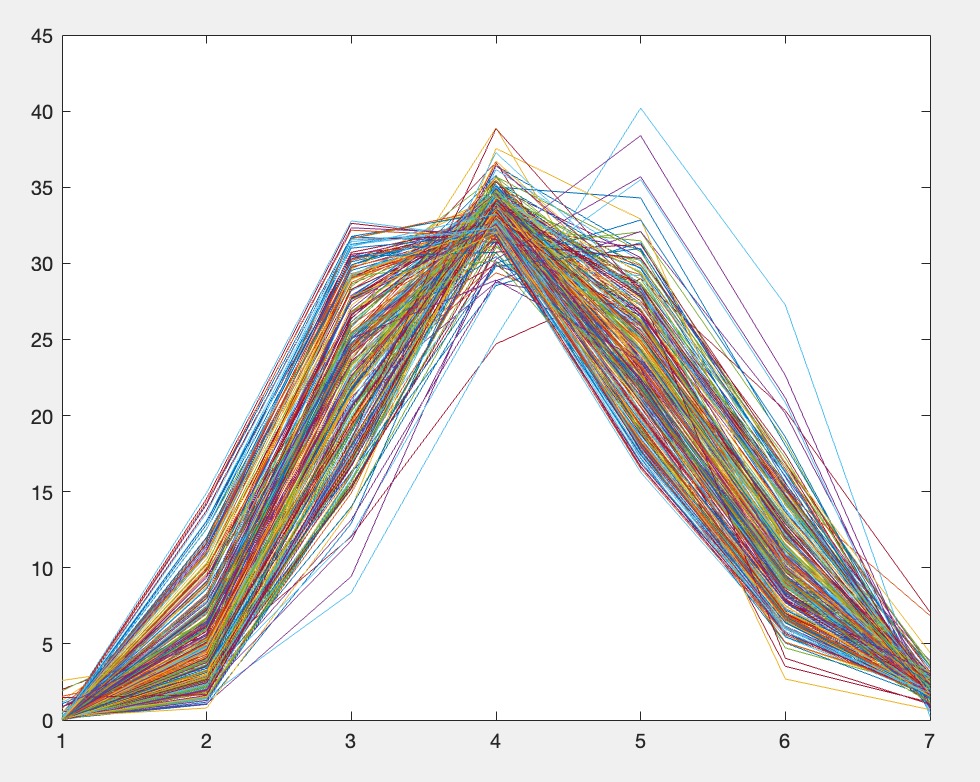}
    \includegraphics[scale = 0.20]{pic13.jpg}\\Figure 11: Comparison of Predicted and Actual Results
\end{center}
After that, we compare our trained model's results with our original testing group. The formula we used to calculate the error was the same in our previous model. That is: $$W = \frac{\sum_{i=1}^{60}(\frac{\sum_{j=1}^7 \frac{D'_j - D_j}{D_j}}{7}) }{60}  $$where:
\begin{itemize}
    \item $j$ is the number of trials players take to guess the word correct. j starts from 1 and ends in 7, indicating the seven different possibilities with 1, 2, 3, 4, 5, 6, and fail(x).
    \item $D'_j$ is the predicted distribution of percentage on the given trial number j.
    \item $D_j$ is the actual distribution of percentage on the given trial number j.
    \item Divided by 60 because our testing group's size is 60.
\end{itemize}
An example of comparison of probability distribution is shown below:
\begin{center}
    \includegraphics[scale = 0.19]{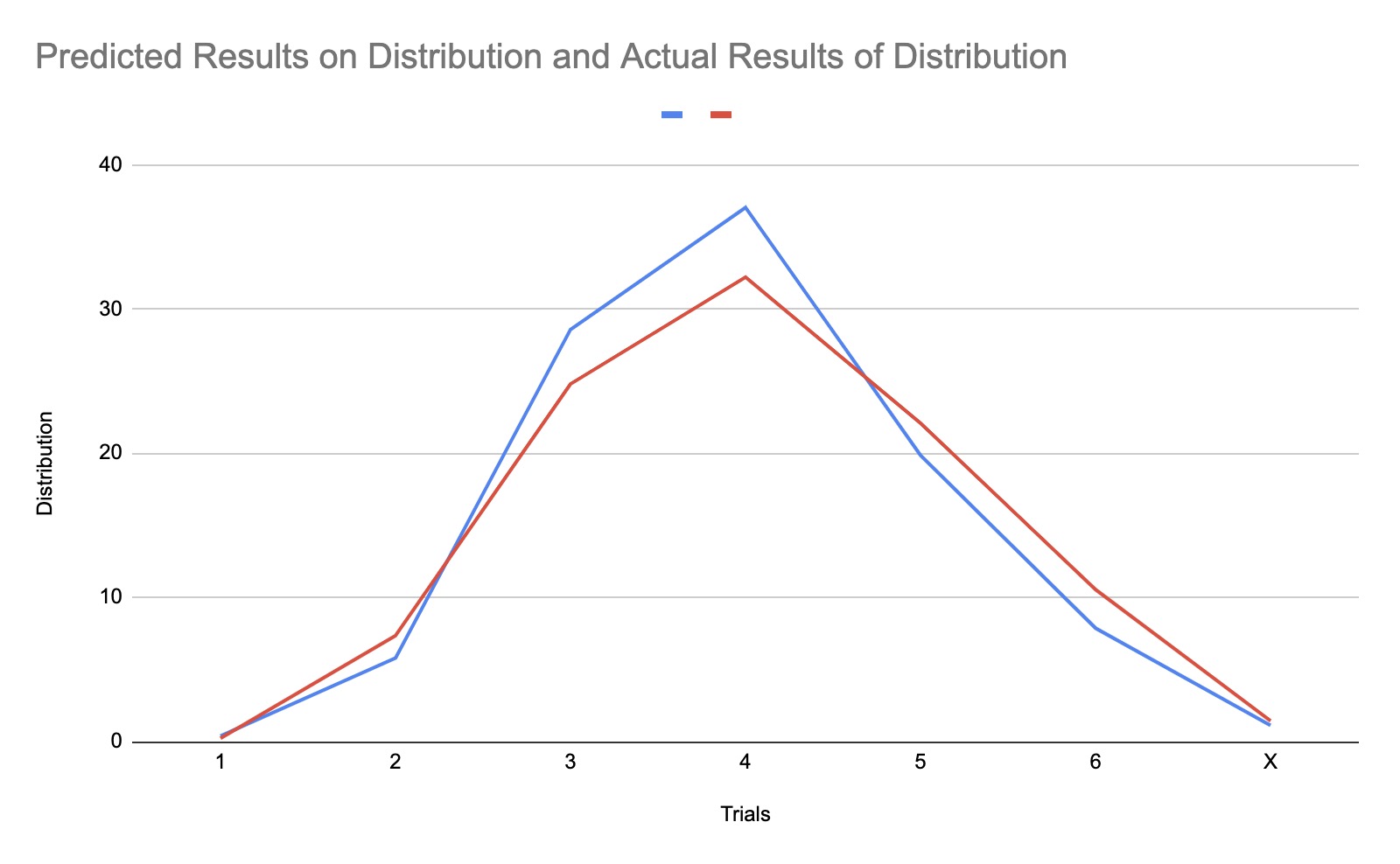}\\Figure 12: An Example of Predicted vs. Actual Results
\end{center}
The analysis of the graph suggests that the predicted quantity is in close proximity to the actual value, however, there is an inherent degree of inaccuracy in the resulting probability distribution. Consequently, the calculated value of $W$ is 0.205968, which corresponds to an error of approximately $21 \%$. Based on our analysis of the data, we observed that the errors for trials 2, 3, 4, 5, and 6 were negligible, with values ranging from 0.001 to 0.02. However, the errors for trials 1 and $x$ were relatively substantial, ranging from 0.1 to 6.0. This discrepancy in error led to a significant increase in the average error of the given word. Two potential explanations for this observation were considered: 1) the data provided may not be accurate due to rounding to whole numbers, and 2) the method utilized for calculating error for small numbers may be unreliable, as the divisor is very small and even minor errors may result in significant deviations. For example, a predicted value of 1.4 with an actual value of 1 would result in an error of $\frac{0.4}{1}$ or $40 \%$. Therefore, this batch of BPNN coefficient will be used in the all 359 days of data in order to predict other word's distribution that we will implement in section 4. 
\subsection{Model For Measuring Difficulty of a Word}
The last model we made is for measuring difficulty of a given word. However, the so called ``difficulty'' is relatively subjective. At the beginning, based on our experience on playing the Wordle game, we decided to simply use the average number of trials it takes for players to solve as an indicator for the difficulty of a given word. This formula is simply calculated by $$D = \sum_{i=1}^7 P_i \cdot i$$where i is the number of trials from 1 to 7, and $P_i$ is the probability of players solved the problem with that trial number. However, we acknowledged that this formula was too basic to predict a given words, since the effect data was only limited with no more than 20,000 players. Therefore, we conducted another model in order to indicate the level of difficulty, which is called K-means clustering algorithm. This is also a machine learning technique that groups similar data points into a predetermined number of clusters based on their distance from each other's centroid. In order to approach this, we used the Machine Learning Toolbox in MATLAB to classify our number of clusters (how many levels of difficulty) as 5, average number of trials as input, and run through the whole dataset through this model. \\
We resulted in using 5 as the number of clusters by employing the Silhouette method on our K-means clusters. By evaluating the Silhouette score of different number of clusters, we have a maximum score at 5 clusters according to the figure below.

\begin{center}
    \includegraphics[scale = 0.22]{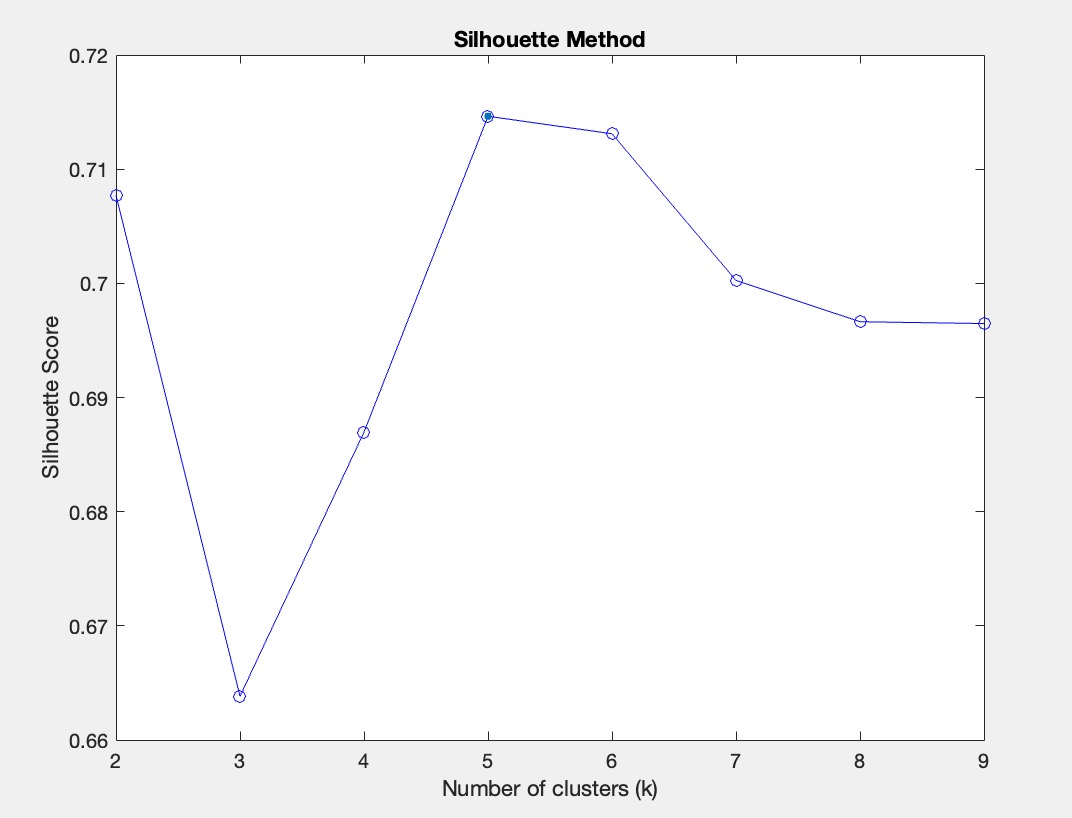} \\ Figure 13: Number of Clusters and Silhouette Scores
\end{center}
As a result, the K-means with 5 clusters indicate that the five different levels will be: 3.59, 3.97, 4.28, 4.59, and 5.00 number of trials. Therefore, we can also classify our words into these five levels and getting the scatter diagram. After sorting them up with Excel, we also obtained the percentage of each difficulty level of our dataset:
\begin{center}
    \includegraphics[scale = 0.19]{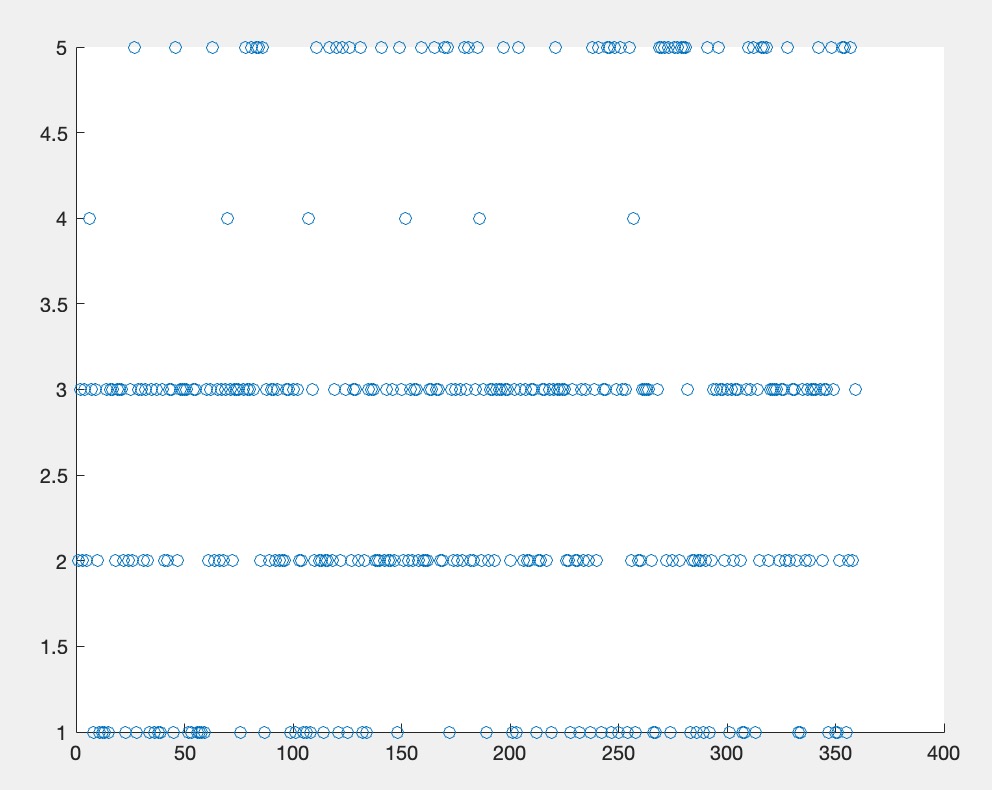} 
    \includegraphics[scale = 0.43]{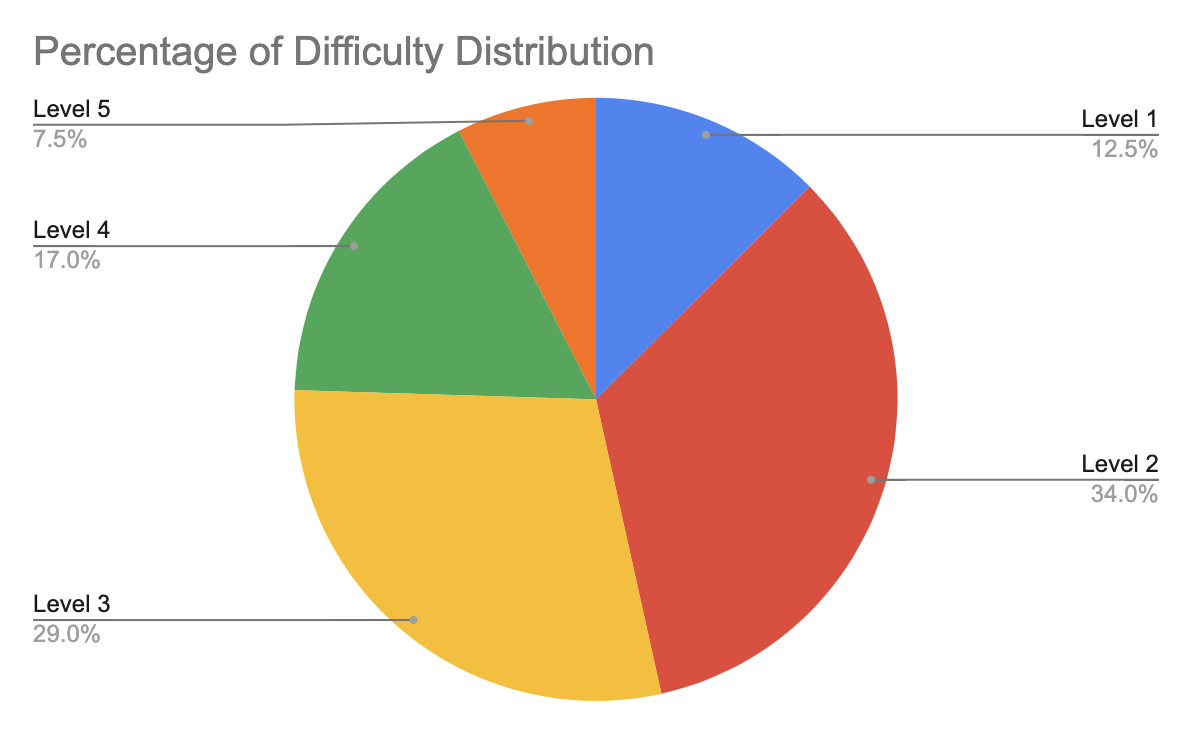} \\ Figure 14: Scale Diagram of Difficulty and Their Distribution
\end{center}
The reason why we used Silhouette Coefficient is because it is one of the most common test for a K-means clustering algorithm. It measures the validity of the algorithm very well. Its formula can be written as: $$s(i) = \frac{b(i) - a(i)}{\max \{a(i), b(i)\}}$$where $i$ is the index of the data point, $a(i)$ is the average distance between the data point $i$ and all other data points in the same cluster, and $b(i)$ is the minimum average distance between the data point $i$ and all other clusters to which $i$ does not belong. As a result, our Silhouette Coefficient is 0.6738. According to its introduction, values closer to 1 indicates that the data point is well-matched to its own cluster and poorly matched to neighboring clusters, and values closer to -1 indicating the opposite. Therefore, our result is precise to a good level and can be used to determine the difficulty. 
\section{Implementation and Results}
\subsection{Number Of Submitted Results Model}
As summarized in section 3.1, we will use our modified ARIMAX model to predict the number of results on March 1st, 2023 with all 281 days (Only the declining period) of data. Simply plug this date into the equation in MATLAB, we obtain a number of 12884 results. No uncertainty was measured during our process of calculation. And the coefficient we calculated in MATLAB is shown below:
\begin{center}
    \includegraphics[scale = 0.24]{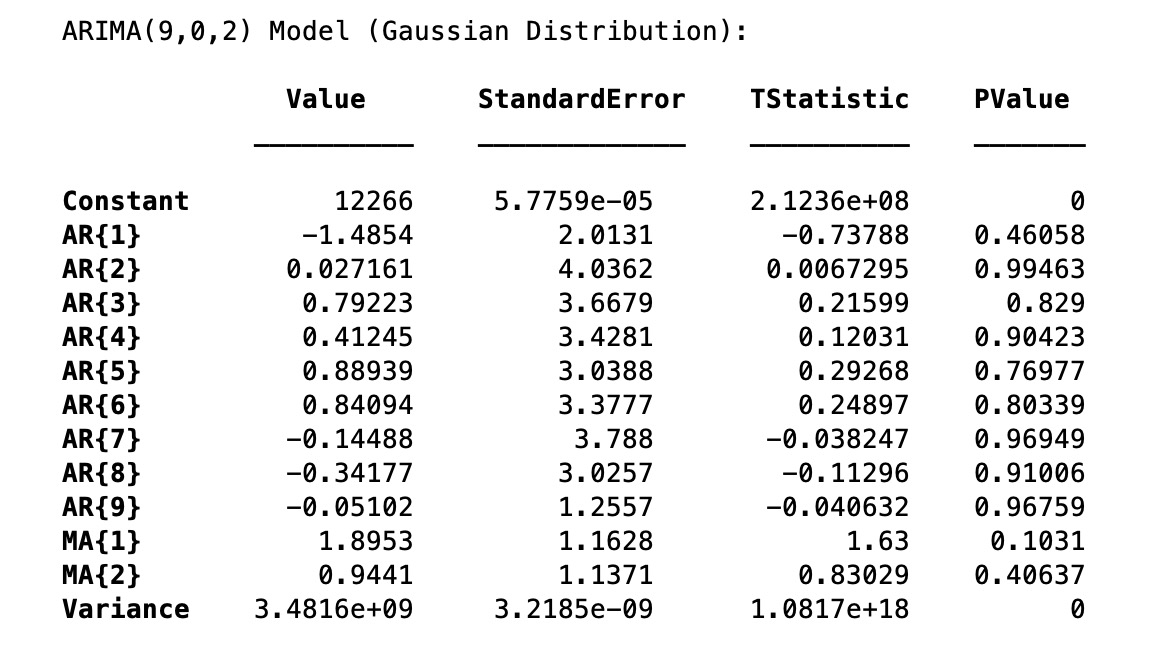} \\Table 5: Coefficients List of Final Model
\end{center}
And the final graph of the date vs. submitted number of results indicated by this ARIMAX model is displayed below, where the blue lines is the reported results and the red line is predicted by our model.
\begin{center}
    \includegraphics[scale = 0.35]{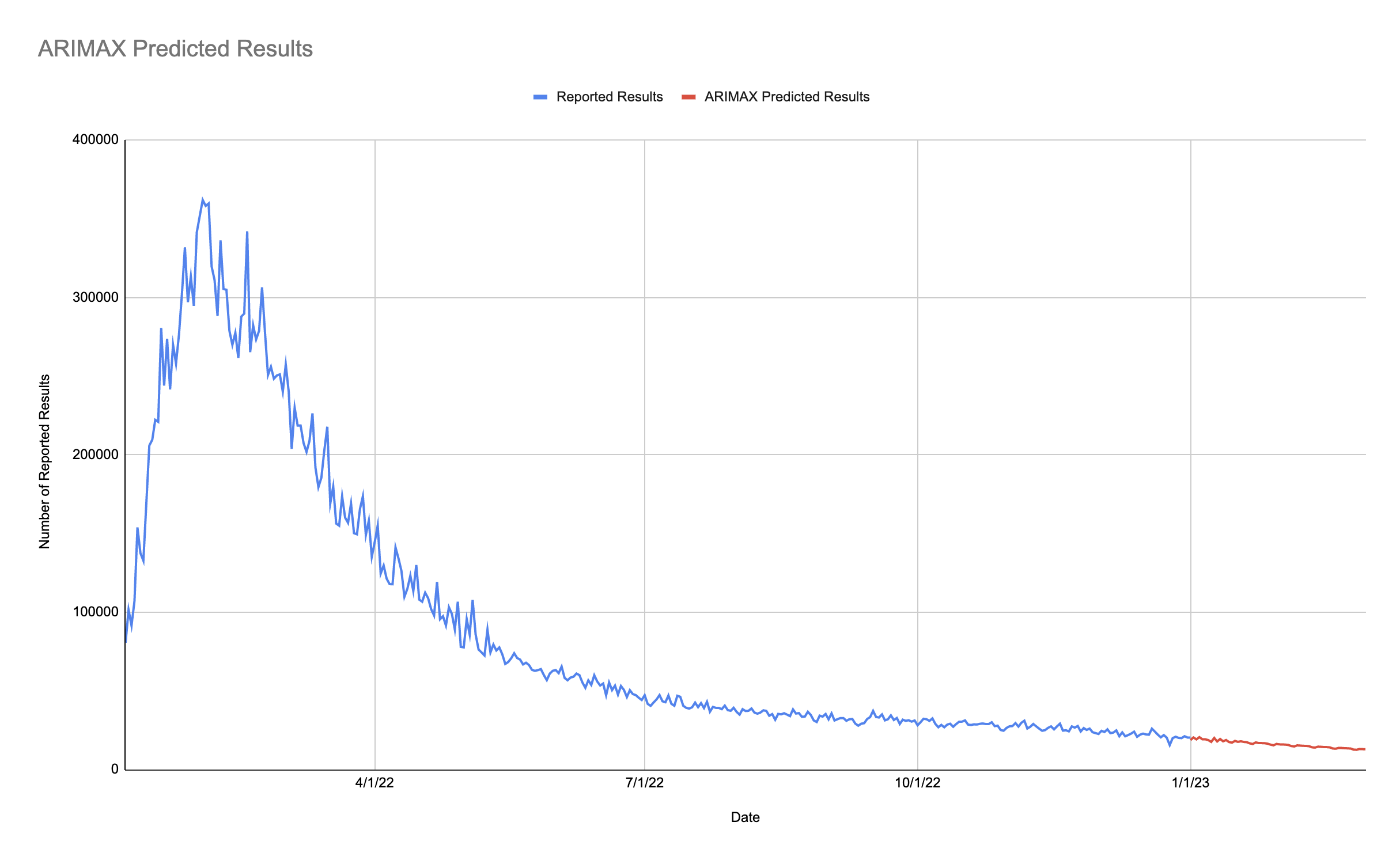}\\Figure 15: Prediction of ARIMAX
\end{center}
\subsection{Distribution vs Word Attributes Model}
Similarly, using our Backpropagation Neural Network model, the input word ``eerie'' on the given date March 1st 2023 (day 419) will generate the result presented in the figure below:
\begin{center}
    \includegraphics[scale = 0.2]{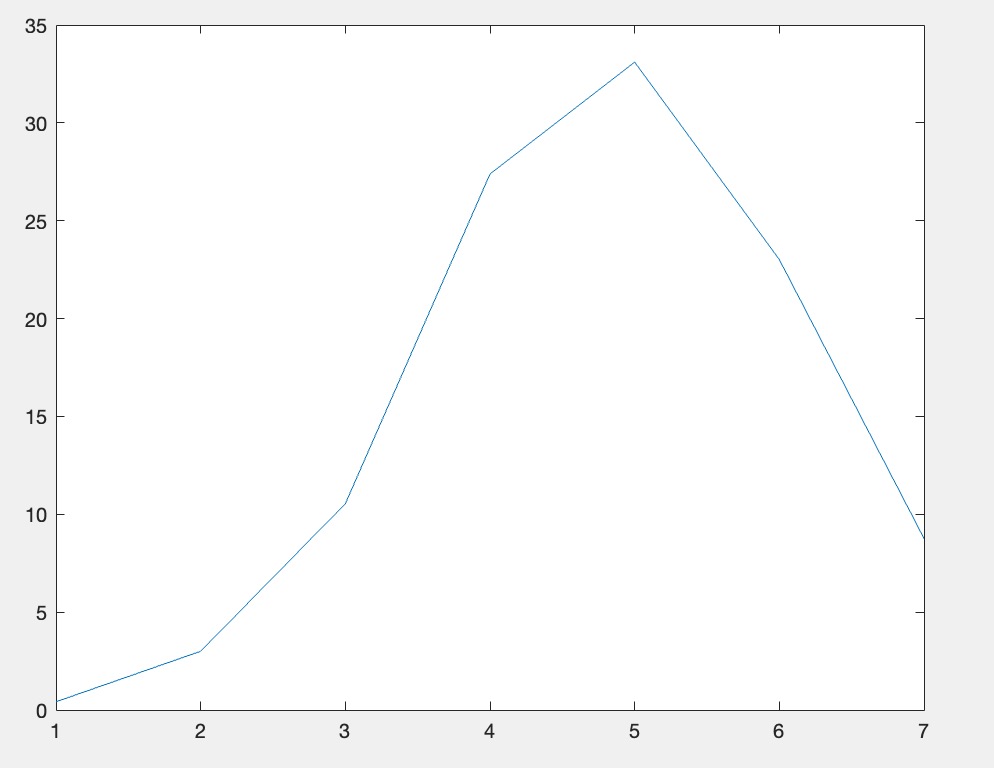}
    \includegraphics[scale = 0.44]{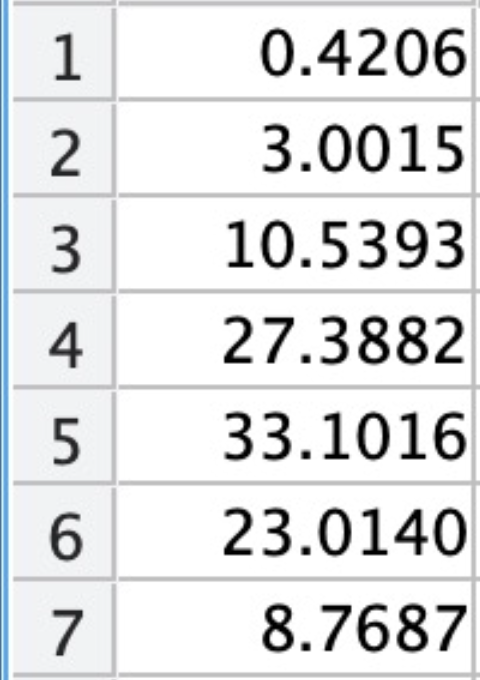} \\Figure 16: Prediction of BPNN on March 1st ``eerie''
\end{center}
As depicted, the percentage distribution is shown in both line graph and grid, where 7 means x, or fail. After that, the average number of trials it take for all players can be calculated by the formula: $$X = \frac{\sum_i^7 i \cdot P_i}{7} $$And as a result, this average number of trials we obtain from the word ``eerie'' is in fact 4.82, which is reasonable because it is indeed a hard word to guess. \\
Therefore, based on our K-means clustering algorithm described in section 3.4, we found that the word ``eerie'' would have an average number of trials $4.82$ Since this number is closer to level 5, 5.00 than level 4, 4.59, we conclude that ``eerie'' is a level 5 difficulty word. \\
To examine the relationship between word attributes and difficulty level, we extracted features for all words within each level (as depicted in Figure 8 and Figure 10). For each level, we computed the average of all words' features, resulting in a list of average features of the level. In the case of character frequency, we utilized the sum of all characters' frequency as an indicator of the occurrence of rare characters. Our analysis identified several word attributes that exhibited correlations with difficulty level, including frequency of characters, number of unique characters, and number of vowels and consonants. The corresponding figures are presented in Figure 17.\\
This analysis reveals a negative correlation between difficulty levels and the number of unique characters as well as the sum of frequencies of characters. This suggests that words with a greater number of repeated and rare letters are more difficult, consistent with the assumption stated in section 3.3.2. Additionally, the number of vowels demonstrates a downward trend as the difficulty level increases, while the number of consonants exhibits an upward trend with increasing difficulty level, which implies that words containing a higher number of vowels may pose greater difficulty for players.
\begin{center}
    \includegraphics[scale = 0.18]{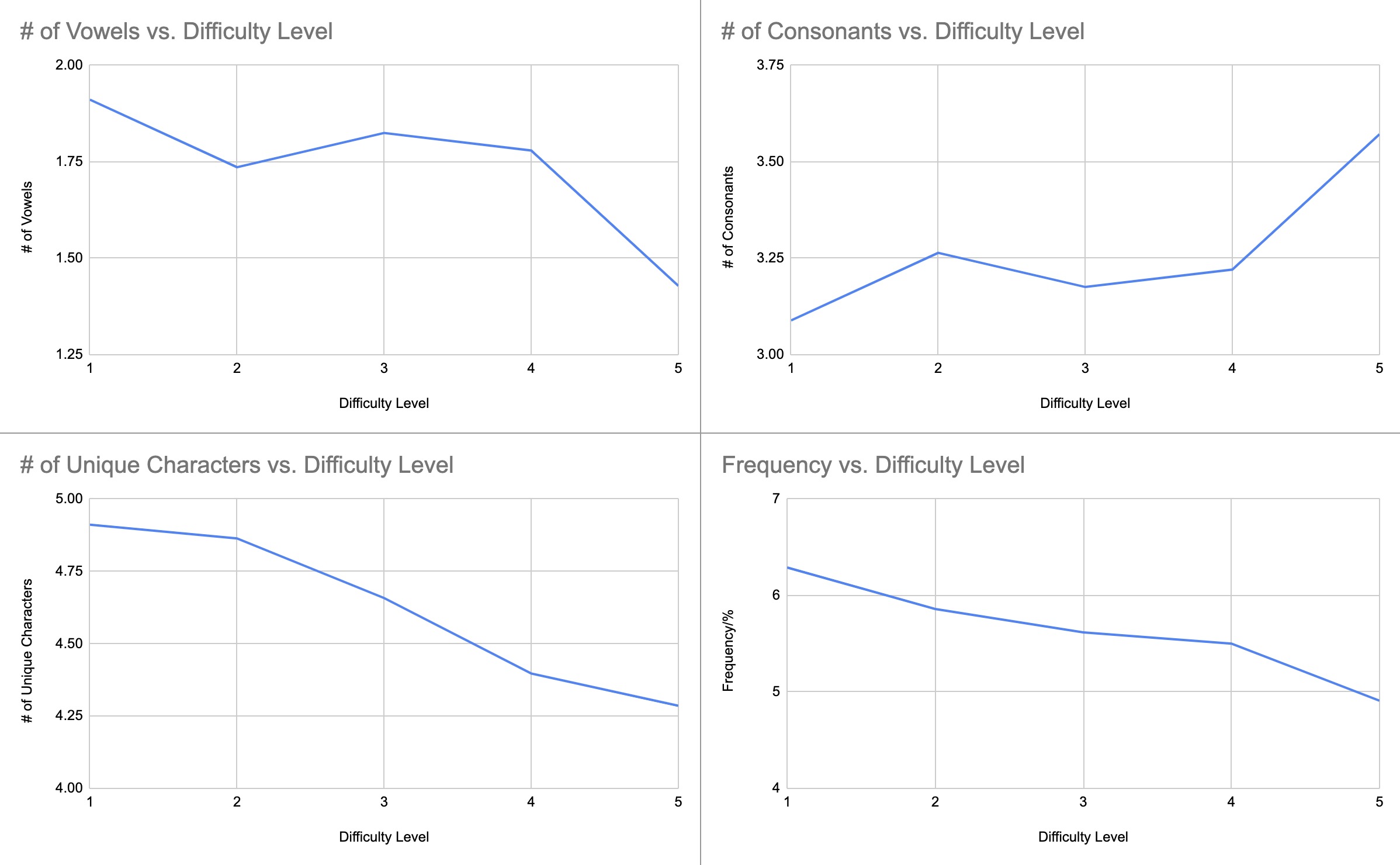} \\Figure 17: Difficulty Level vs. Word Attributes
\end{center}
\section{Validation and Sensitivity Analysis}
To perform a comprehensive and accurate validation analysis, we employed the mean absolute error (MAE) as evaluation metrics for our models \cite{DIGITAL_DESIGN_AND_COMPUTER_ARCHITECTURE}. For our machine learning model, we applied cross-validation techniques to validate its precision. \\
First of all, we applied MAE test to our ARIMAX model with the formula described below: $$MAE = \frac{\sum_{i=1}^n |X'_i - X_i|}{n}$$where $X'_i$ is the predicted value in day i and $X$ is the actual value. Using the same method in section 3, we still implement our model on the training group then followed with finding the MAE in the testing group. Based on our calculation through MATLAB, the MAE for our ARIMAX model is 664.643. Since the data for this problem is all 5-digit numbers, we consider our estimation relatively accurate. For our BPNN model that predicts the probability distribution, we conducted a cross validation, which splits data into other subsets of training and testing that differed from the original sets. Therefore, instead of the training set being the first 300 words, we flipped the order where the training group is the last 300 days of words, than test the validation on the first 59 days of word. The newly trained model is depicted below:
\begin{center}
    \includegraphics[scale = 0.36]{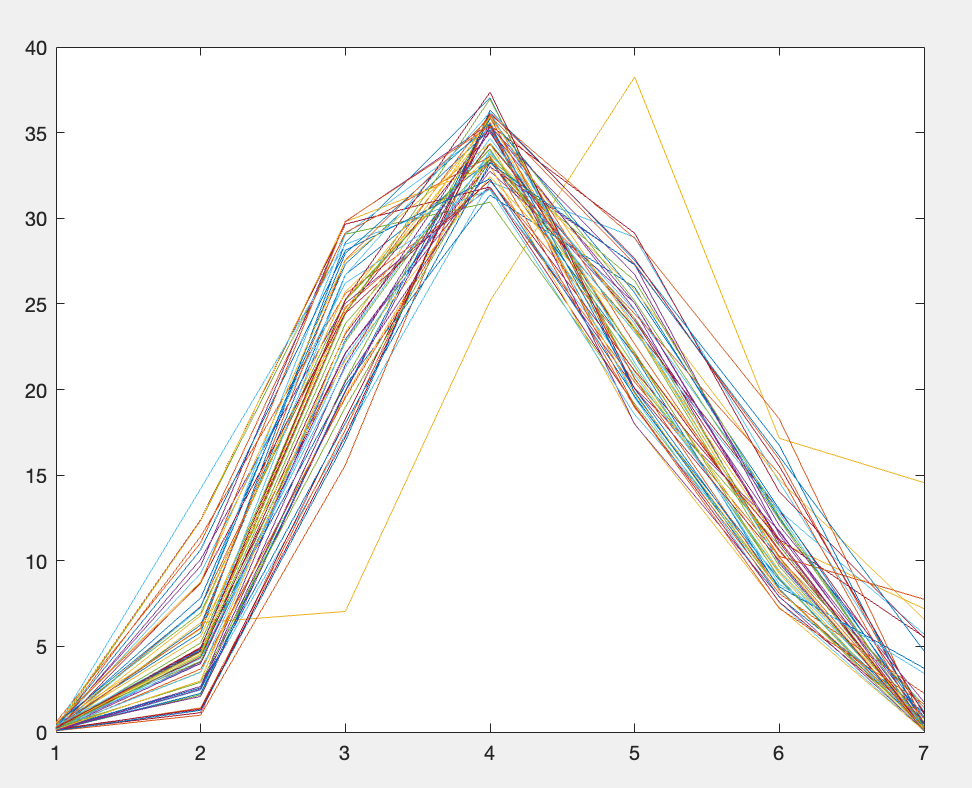}
    \includegraphics[scale = 0.36]{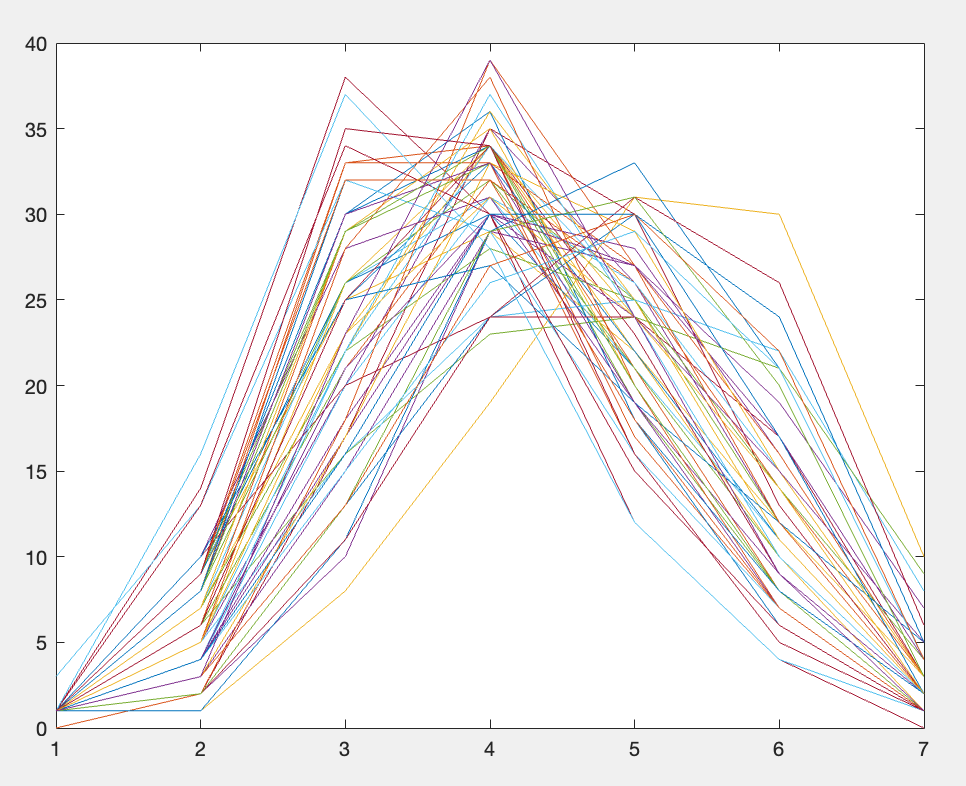} \\Figure 18: Cross Validation Results
\end{center}
The left hand side is our trained model predicting on the new testing group (day 1 - day 59). The right hand side is the actual results. It can be seen that although the model has some level of misfit, the graph still obtain a general shape accordingly. Using the same method for calculating errors we adopted in section 3, we found that our error for this cross validating group is 0.3422, which is considerably similar to the original model. Besides model validation, we also conducted a sensitivity analysis for our models. As for this particular problem, we have conceived the following scenario: after the COMAP released the problem, the ICM/MCM participants that chose to work on problem C all over the world hopped into New York Times website and attempted to solve that day's Wordle problem by themselves. According to the data posted in the official COMAP website, 95$\%$ of ICM teams came from China, a non-English speaking country. In 2022, around 10,000 teams chose problem C, which equivalent to at most $30,000 * 95\% = 28500$ non-English speaking people. Therefore, If I add these numbers to our model from Feb 16 to Feb 18 and let their score normally distributes in trial 4, 5, 6, x (since most of them do have have a good vocabulary), the new graph on February is depicted below after the sudden change in this date:
\begin{center}
    \includegraphics[scale = 0.36]{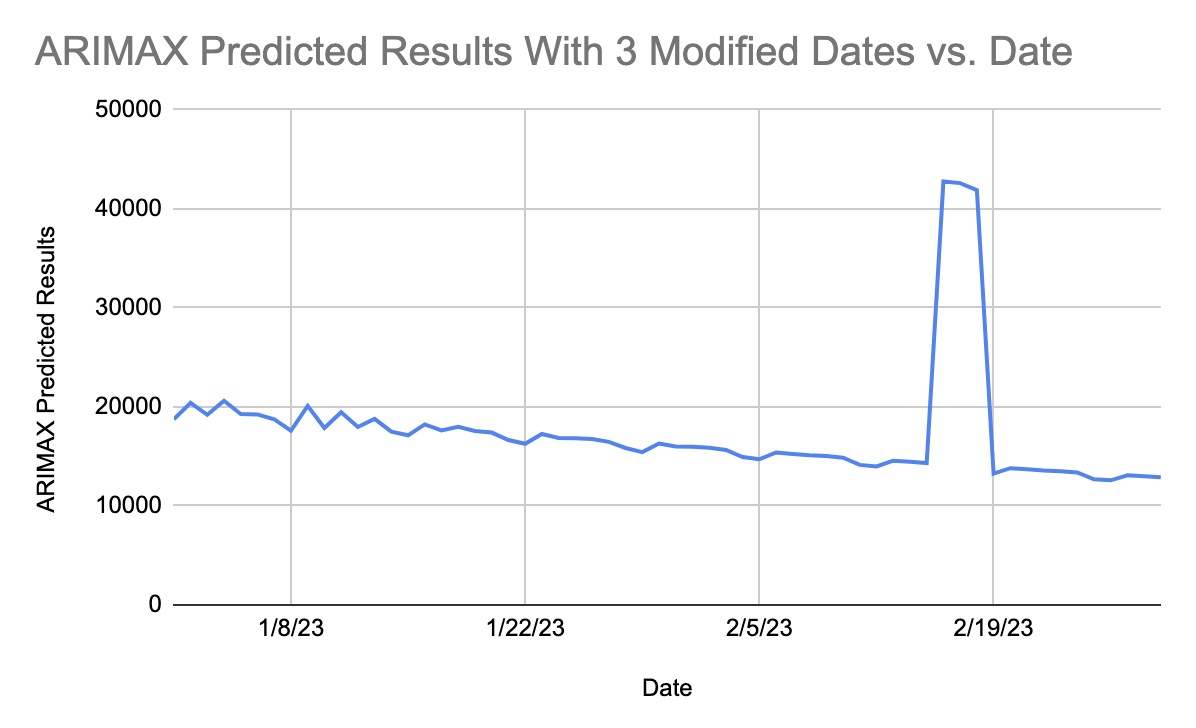} \\Figure 19: Sensitivity Analysis Input
\end{center}
After that, we run through the previous 359 days plus this modified dates into our ARIMAX model. The new predicted results on March 1 based on the previous 371 days is 12194 submitted results, with only $5.35\%$ error. This percentage of error is small enough in our expectation, indicating our ARIMAX(9, 0, 2) has a good level of robustness. 
\section{Interesting Features of The Dataset}
During our processing of writing the paper, we found some interesting aspects of this data. First, the hard mode percentage starts with a very little amount, then increases and tends to be stabilized around $8\%$. Similarly, the number of submitted results increases to a great amount, then decreases and tends to be stabilized around 20,000. As mentioned in previous sections, our team suggested that it was because the promotion made by New York Times, the number of players suddenly increased at the beginning two months of the year. After the heat of the promotion, people start to lose interest gradually so that the remaining players are the real dedicated fans of Wordle. Among those players, about $8\%$ of them are people who'd willing to take challenges, or play in the hard mode. 

\section{Conclusion}
In this research paper, we present the results of our analysis of the Wordle game. Our study employs an ARIMAX model, which predicts that the number of results submitted on March 1st, 2023 will be 12884. Additionally, our BPNN model indicates that the word ``eerie'' will result in an average of 4.82 trials on that day. We also use a K-means clustering algorithm to determine the difficulty level of the word, which is found to be level 5, the highest level. The clustering algorithm identifies the association of the difficulty level 
and attributes of a word, such as total letter frequency, number of unique characters, vowel counts, and consonant counts. \\
One limitation of our study is that we initially identified the percentage of words distributed across the hard mode without realizing that this applied to all reported results. However, a strength of our approach is that we use a variety of techniques and tools, including machine learning, human observation, continuous data, discrete data, regression, cross-validation, data bucketing through pie charts, R, and MATLAB. We acknowledge that our models may still contain some errors that cannot be entirely eliminated. We also notice that some of our observations, such as the Consecutive Uncommon Letters could not be put into the model with the relatively short amount of time we had.\\
In future research, we aim to collect and record new data for every day's Wordle game and refine our machine learning model to further reduce our error. Overall, our study provides insights into the Wordle game and demonstrates the utility of machine learning in analyzing language data. Future research could also dive into how Consecutive Uncommon letters affect the guesses made.
\pagebreak

\end{document}